# Multi-Layered Blockchain Governance Game

SONG-KYOO (AMANG) KIM

## ABSTRACT

The research designs a new integrated system for security enhancement of a decentralized network by preventing damages from attackers particularly for the 51 percent attack. The concept of multiple layered design based on Blockchain Governance Games frameworks could handle multiple number of networks analytically. The Multi-Layered Blockchain Governance Game is an innovative analytical model to find the best strategies for executing a safety operation to protect whole multiple layered network systems from attackers. This research fully analyzes a complex network with the compact mathematical forms and theoretically tractable results for predicting the moment of a safety operation execution are fully obtained. Additionally, simulation results are demonstrated to obtain the optimal values of configuring parameters of a blockchain based security network. The Matlab codes for the simulations are publicly available to help those whom are constructing an enhanced decentralized security network architecture through this proposed integrated theoretical framework.

## 1. INTRODUCTION

Recently, the innovative security mechanisms by adapting the blockchain and the game theory have been designed by the author [8, 9]. The *Blockchain Governance Game* (BGG) which has firstly proposed by the author is the stochastic game model with the fluctuation and the mixed strategy game for analyzing the network to provide the decision making moment for taking preliminary security actions before attacks [8]. The model is targeted to prevent blockchain based attacks (i.e., the 51 percent attack) and keeps the network decentralized. The *Strategic Alliance for Blockchain Governance Game* (SABGG) is designed for allying the nodes instead of keeping honest nodes by a defender. In the business, the strategic alliance is an agreement within two or more parties to pursue a set of agreed upon objectives needed and it has emerged to solve many company business problems [9]. The governance in the Blockchain is followed by the decision making parameter which are includes the prior time just before an attacker catches more than half of the total nodes. We will not take any action until the time when it passes the first passage time and it still have the chance that all nodes are governed by an attacker even an attacker catches less than the half of nodes. The Multi-layer Blockchain Governance Game (MLBGG) is a combined stochastic model with the sets of the BGG and SABGG models (see Figure 1). The layer 1 is a set of multiple BGG based networks and the other layer (i.e., layer 0) is single SABGG based network.

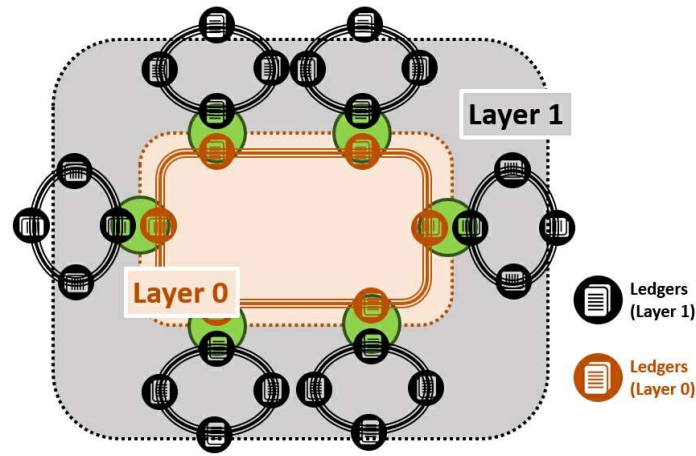

**Figure 1.** Multi-layer Blockchain Governance Game structure

This multiple layer framework makes the MLBGG to be adapted into various hierarchical system architectures including Edge-Fog computing [10], hierarchical network systems [11] and IoT-Server networks (see Section 3).

## 2. MULTI-LAYERED BLOCKCHAIN GOVERNANCE GAME

A multi-layered blockchain governance game (MLBGG) combines a set of multiple BGG networks and single SABGG network which are hierarchically connected. Each system in the layer 1 is exactly mapped with the BGG and the layer 0 is mapped with the SABGG (see Figure 2).

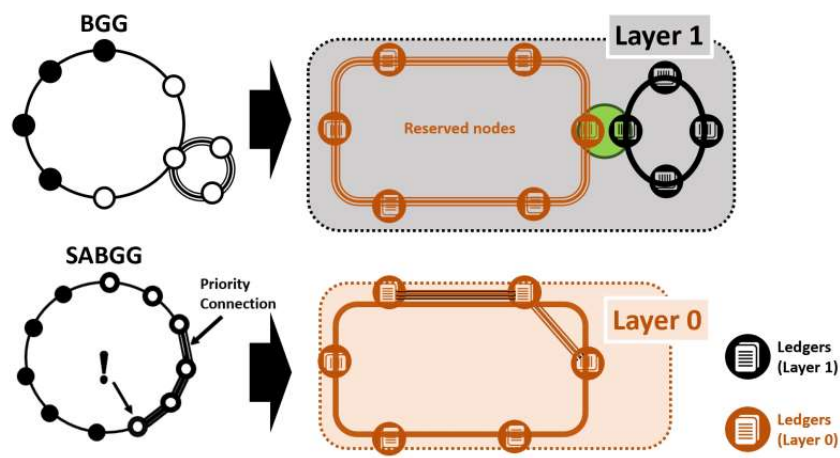

**Figure 2.** BGG and SABGG mapping into MLBGG

## 2.1. Layer-1 Basic Stochastic Model

In $l$-th blockchain network of the layer 1 (i.e., the second layer), the antagonistic game of two players (called "A" and "H") are introduced to describe the blockchain network between a defender and an attacker. Both players compete to build the blocks either for honest or false ones. Let $(\Omega, \mathcal{F}(\Omega), P)$ be probability space $\mathcal{F}_A$, $\mathcal{F}_H$, $\mathcal{F}_\tau \subseteq \mathcal{F}(\Omega)$ be independent $\sigma$-subalgebras in the $l$-th layer. Suppose:

$$\mathcal{A}^l := \sum_{k \geq 0} X_k^l \varepsilon_{s_k^l}, \quad s_0^l(=0) < s_1^l < s_2^l < \cdots, \text{ a.s.} \tag{2.1}$$

$$\mathcal{H}^l := \sum_{j \geq 0} Y_j^l \varepsilon_{t_j^l}, \quad t_0^l(=0) < t_1^l < t_2^l < \cdots, \text{ a.s.} \tag{2.2}$$

are $\mathcal{F}_A$-measurable and $\mathcal{F}_H$-measurable marked Poisson processes ($\varepsilon_a$ is a point mass at $a$) with respective intensities $\lambda_A^l$ and $\lambda_H^l$ and point independent marking. The game in this case is a stochastic process $\mathcal{A}_\tau^l \otimes \mathcal{H}_\tau^l$ describing the evolution of a conflict between players A and H known to an observation process. The game is over when on the $k$th observation epoch $\tau_k$, the collateral building blocks to player A exceeds more than the half of the total nodes $M$ of each blockchain network in the layer 1. To further formalize the game, the *exit index* is introduced:

$$\nu^l := \inf\{k : A_k^l = A_0^l + X_1^l + \cdots + X_k^l \geq \left(\tfrac{M_l}{2}\right)\}, \tag{2.15}$$

$$\mu^l := \inf\{j : H_j^l = H_0^l + Y_1^l + \cdots + Y_j^l \geq \left(\tfrac{M_l}{2}\right)\}. \tag{2.16}$$

Since, an attacker is win at time $\tau_\nu$, otherwise an honest node generates the correct blocks. We shall be targeting the confined game in the view point of player A. The first passage time $\tau_\nu$ is the associated exit time from the confined game and the formula (2.6) will be modified as

$$\overline{\mathcal{A}_\tau^l} \otimes \overline{\mathcal{H}_\tau^l} := \sum_{k \geq 0}^{\nu^l} (X_k^l, Y_k^l) \varepsilon_{\tau_k} \tag{2.17}$$

which the path of the game from $\mathcal{F}(\Omega) \cap \{\nu^l < \mu^l\}$, which gives an exact definition of the model observed until $\tau_{\nu^l}$. The joint functional of the blockchain network model is as follows:

(2.18)
$$\Phi^l_{\left\lceil \frac{M_l}{2} \right\rceil} = \Phi^l_{\left\lceil \frac{M_l}{2} \right\rceil}(\xi, g_0, g_1, z_0, z_1)$$
$$= \mathbb{E}\left[\xi^{\nu^l} \cdot g_0^{A_{\nu^l-1}^l} \cdot g_1^{A_{\nu^l}^l} \cdot z_0^{H_{\nu^l-1}^l} \cdot z_1^{H_{\nu^l}^l} \mathbf{1}_{\{\nu^l < \mu^l\}}\right], \quad l = 0, \ldots, N,$$

where $M_l$ indicates the total number of nodes (or ledgers) of the $l$-th blockchain network in the layer 1. This functional will represent the status of attackers and honest nodes upon the exit time $\tau_\nu^l$. The latter is of particular interest, we are interested in not only the prediction of catching up the blocks by attackers but also one observation prior to this. The Theorem 1 (BGG-1) establishes in [8] and the first exceed model by Dshahalow [12] has been adapted and its operators are defined as follows:

$$\mathcal{D}_{(x,y)}\left[f(x,y)\right](u,v) := (1-u)(1-v)\sum_{x\geq 0}\sum_{y\geq 0}f(x,y)u^x v^y, \tag{2.19}$$

where $\{f(x,y)\}$ is a sequence, with the inverse

$$\mathfrak{D}_{(u,v)}^{(m,n)}(\bullet) = \begin{cases} \left(\frac{1}{m!\cdot n!}\right)\lim_{(u,v)\to 0}\frac{\partial^m \partial^n}{\partial u^m \partial v^n}\frac{1}{(1-u)(1-v)}(\bullet), & m\geq 0, n\geq 0, \\ 0, & \text{otherwise.} \end{cases} \tag{2.20}$$

Additionally, the new operators for dealing the matrix calculations are introduced. Let us consider the matrix of a function set $\boldsymbol{f}_{(x,y)}$ as follows:

$$\boldsymbol{f}_{(x,y)} = \begin{bmatrix} f_1(x,y) \\ f_2(x,y) \\ \vdots \\ f_l(x,y) \\ \vdots \end{bmatrix} \tag{2.21}$$

and the matrix operations for $\mathcal{D}_{(x,y)}\{\bullet\}$ and $\mathfrak{D}_{(u,v)}^{(m,n)}\{\bullet\}$ are defined as follows:

$$\mathbb{D}\odot\boldsymbol{f}_{(x,y)} := \mathcal{D}_{(x,y)}\{\boldsymbol{f}\} = \begin{bmatrix} (1-u)(1-v)\sum\sum f_1(x,y)u^x v^y \\ (1-u)(1-v)\sum\sum f_2(x,y)u^x v^y \\ \vdots \\ (1-u)(1-v)\sum\sum f_l(x,y)u^x v^y \\ \vdots \end{bmatrix} \tag{2.22}$$

and

$$\mathbb{D}_M^{-1}\odot\boldsymbol{G}_{(u,v)} := \mathfrak{D}_{(u,v)}^{(m,n)}\left(\begin{bmatrix} G_1(u,v) \\ G_2(u,v) \\ \vdots \\ G_l(u,v) \\ \vdots \end{bmatrix}\right) = \begin{bmatrix} \mathfrak{D}_{(u,v)}^{(m_1,n_1)}\{G_1(u,v)\} \\ \mathfrak{D}_{(u,v)}^{(m_2,n_2)}\{G_2(u,v)\} \\ \vdots \\ \mathfrak{D}_{(u,v)}^{(m_l,n_l)}\{G_l(u,v)\} \\ \vdots \end{bmatrix}, \tag{2.23}$$

where

$$\boldsymbol{M} = \begin{bmatrix} m_1 & n_1 \\ m_2 & n_2 \\ \vdots & \vdots \\ m_l & n_l \\ \vdots & \vdots \end{bmatrix}. \tag{2.24}$$

From (2.18) and (2.27), we can find the PGFs (probability generating functions) of $A_{\nu-1}^l$ (and $A_\nu^l$) and the *exit index* of each blockchain network in the layer 1.

$$\mathbb{E}\left[\xi^{\nu^l}\right] = \Phi_{\left\lceil\frac{M_l}{2}\right\rceil}^l(\xi,1,1,1,1), \tag{2.34}$$

$$\mathbb{E}\left[g_0^{A_{\nu-1}^l}\right] = \Phi_{\left\lceil\frac{M_l}{2}\right\rceil}^l(1,g_0,1,1,1), \tag{2.35}$$

$$\mathbb{E}\left[g_1^{A_\nu^l}\right] = \Phi_{\left\lceil\frac{M_l}{2}\right\rceil}^l(1,1,g_1,1,1), \quad l = 0,\ldots,N. \tag{2.36}$$

The moment of making a decision $\tau_{\nu-1}$ could be found from (2.4), (2.10) and (2.34):

$$\mathbb{E}\left[\nu^l\right] = \left.\frac{\partial}{\partial \xi}\Phi_{\left\lceil\frac{M}{2}\right\rceil}^l(\xi,1,1,1,1)\right|_{\xi=1}, \tag{2.37}$$

$$\mathbb{E}\left[\tau_{\nu^l-1}^l\right] = \mathbb{E}\left[\tau_0^l\right] + \mathbb{E}\left[\Delta_1^l\right]\left(\mathbb{E}\left[\nu^l\right]-1\right), \quad l = 1,\ldots,n. \tag{2.38}$$

In a conventional BGG, the probability of bursting the $l$-th blockchain network $q^l(s_H)$ is determined as follows:

$$q^l(s_H) = \begin{cases} \mathbb{E}\left[\mathbf{1}_{\{A_\nu^l \geq \frac{M}{2}\}}\right], & s_H = \{DoNothing\}, \\ \mathbb{E}\left[\mathbf{1}_{\{A_\nu^l \geq \left(\frac{M}{2}+B\right)\}}\right], & s_H = \{Action\} \end{cases} \tag{2.39}$$

where $B(\leq \eta)$ is the number of reserved nodes which will be joined in the network. It is noted that the reserved nodes depend on the availability of other blockchain networks which are hooked as the layer 0 (see Figure 2). Hence, the bursting probability when the reserved nodes are realized during a safety mode is revised as follows:

$$\mathbb{E}\left[\mathbf{1}_{\{A_\nu^l \geq \left(\frac{M_l}{2}+B\right)\}}\right] = \mathbb{E}\left[\mathbb{E}\left[\mathbf{1}_{\{A_\nu \geq \left(\frac{M_l}{2}+B\right)\}}\big|B\right]\right] \tag{2.40}$$

where

$$P\{B_\eta = j\} = \binom{\eta}{j}(\rho^1)^j(1-\rho^1)^{\eta-j}, \tag{2.41}$$

$$\rho^1 = \left(\frac{1}{\eta+1}\right)\mathbb{E}\left[\sum_{k=0}^\eta \mathbf{1}_{\{H_\nu^k \geq \frac{M_k}{2}\}}\right], \quad k = 0,\ldots,\eta. \tag{2.42}$$

### 2.2. Layer-0 Basic Stochastic Model

In the Blockchain network of the layer 0 (i.e., the second layer), the antagonistic game of two players (called "Corrupted" and "Genuine") is set up and players compete to build the blocks which are either genuine (or honest) or corrupted (or false). Let $(\Omega, \mathcal{F}(\Omega), P)$ be a probability space and $\mathcal{F}_c, \mathcal{F}_g, \mathcal{F}_t \subseteq \mathcal{F}(\Omega)$ be independent $\sigma$-subalgebras. Suppose:

$$\mathcal{C} := \sum_{j \geq 0} J_j \varepsilon_{u_j}, \quad u_0(=0) < u_1 < u_2 < \cdots, \text{ a.s.} \tag{2.43}$$

$$\mathcal{G} := \sum_{k \geq 0} K_k \varepsilon_{v_k}, \quad v_0(=0) < v_1 < v_2 < \cdots, \text{ a.s.} \tag{2.44}$$

are $\mathcal{F}c$-measurable and $\mathcal{F}_g$-measurable marked Poisson processes ($\varepsilon_w$ is a point mass at $w$) and position independent marking with respective intensities $\lambda_c$ and $\lambda_g$. Before specifying (2.43) and (2.44), a third-party observation point process [12] is equivalent to the duration of the Proof-of-Work (or the Proof-of-Sake) completion [1, 8, 9]:

$$\mathcal{U} := \sum_{i \geq 0} \varepsilon_{t_i}, \ t_0(\ > 0)), \ t_1, \ldots. \tag{2.45}$$

To further formalize the game, the *exit indexes* are defined as follows:

$$\nu := \inf\{j : C_j \ ( = C_0 + J_1 + \cdots + J_j) \geq \left(\tfrac{\eta}{2}\right)\}, \tag{2.58}$$
$$\nu_2 := \inf\{j : C_j \ ( = C_0 + J_1 + \cdots + J_j) - B \geq \left(\tfrac{\eta}{2}\right)\}, \tag{2.59}$$
$$\mu := \inf\{l : G_l ( = G_0 + K_1 + \cdots + K_l) \geq \left(\tfrac{\eta}{2}\right)\}, \tag{2.60}$$

where $B$ is the number of available nodes in the network system (i.e., $B \leq \eta$) by the strategic alliance. Player G might be defeated at $t_\nu$ without requesting the alliance but player G can only be defeated at $t_{\nu_2}$ when the allied genuine nodes are supported. With the strategic alliance, player C could only win the game at time $t_{\nu_2}$ otherwise player G might beat player C at $t_\mu$ when it takes the place beforehand. The game is over at $min\{\nu, \nu_2, \mu_1, \mu\}$ and the $\sigma$-subalgebra of the process $(\mathcal{C}, \mathcal{G})$ can be analogously as $\mathcal{F}(\Omega) \cap \{\nu < \nu_2 < \mu\}$ (i.e., player C wins first). We shall be targeting the confined game of player C. The first passage time $t_\nu$ is the associated exit time from the confined game and the formula (2.51) is modified as

$$\overline{\mathcal{C}_t} \otimes \overline{\mathcal{G}_t} := \sum_{n \geq 0}^{\nu} (J_n, K_n)\varepsilon_{t_n}, \tag{2.61}$$

which gives an exact definition of the model observed until $t_\nu$ without the strategic alliance action. The joint functional of the Blockchain network model with the strategic alliance is as follows:

$$\Theta_{\frac{\eta}{2}} = \Theta_{\{\nu,\nu_2,\mu\}}(\zeta, y_0, y_1, b, z_0, z_1) \tag{2.62}$$
$$= \mathbb{E}\left[\zeta^\nu \cdot y_0^{C_{\nu-1}} \cdot y_1^{C_\nu} \cdot b^{C_\nu - B_\eta} \cdot z_0^{G_{\mu-1}} \cdot z_1^{G_\mu} \mathbf{1}_{\{\nu < \nu_2 < \mu\}}\right]$$

where $\eta$ indicates the total number of nodes (or ledgers) in a Blockchain network. The moment of making a decision $t_{\nu-1}$ could be found from (2.46), (2.54) and (2.77):

$$\mathbb{E}[\nu] = \left.\tfrac{\partial}{\partial \zeta}\Theta_{\left[\frac{\eta}{2}\right]}(\zeta, 1, 1, 1, 1, 1)\right|_{\zeta = 1}, \tag{2.81}$$

$$\mathbb{E}[t_{\nu-1}] = \mathbb{E}[t_0] + \mathbb{E}[U_1](\mathbb{E}[\nu] - 1). \tag{2.82}$$

In the layer 0, the probability of bursting blockchain network $q^0(s_H)$ is determined as follows:

$$q^0(s_g) = \begin{cases} \mathbb{E}\left[\mathbf{1}_{\{C_\nu \geq \frac{\eta}{2}\}}\right], & s_g = \{DoNothing\}, \\ \mathbb{E}\left[\mathbf{1}_{\{C_\nu \geq \frac{\eta(1+\alpha)}{2}\}}\right], & s_g = \{Action\}. \end{cases} \tag{2.83}$$

where $\alpha$ is an overhead for protecting the layer 0 blockchain network on the top of the networks in the layer 1. The probability of bursting a Blockchain network (i.e., an

attacker wins the game) under the memoryless properties becomes the Poisson compound process:

$$q(s_g) = \begin{cases} \sum_{k > \frac{\eta}{2}} \mathbb{E}\big[\mathbf{1}_{\{C_\nu = k\}}\big], & s_g = \{DoNothing\}, \\ \mathbb{E}\left[\sum_{k > \frac{\eta(1+\alpha)}{2}} \mathbb{E}\big[\mathbf{1}_{\{C_\nu = k\}}\big]\right], & s_g = \{Action\}, \end{cases} \quad (2.84)$$

where

$$\mathbb{E}\big[\mathbf{1}_{\{C_\nu = k\}}\big] = \mathbb{E}\left[\mathbb{E}\left[\frac{(\lambda_c t_\nu)^k}{k!} \cdot e^{-\lambda_c t_\nu} \bigg| t_\nu\right]\right]. \quad (2.85)$$

# 3. MULTI LAYERED IOT-SERVER NETWORK DESIGN

## 3.1. Multi-layered IoT-Server Network Architecture

The networks which combines IoT networks and a server network could adapt the MLBGG. The set of IoT networks is one layer and one set with management servers is the other layer. The BGG is adapted into IoT networks as the layer 1 and the SABGG is applied into the network of management servers (see Figure 3).

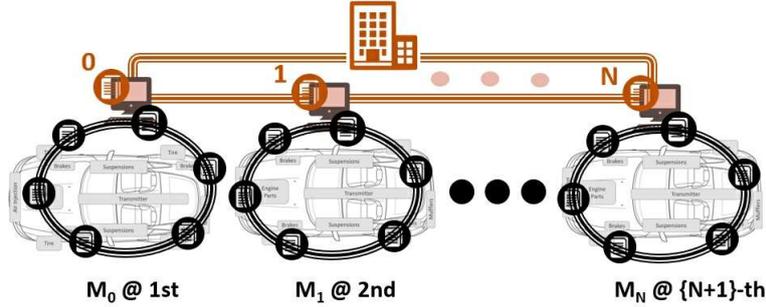

**Figure 3.** IoT-Server network architecture in the MLBGG

## 3.2. Stochastic Optimization

It is noted that the cost for the reserved nodes (i.e., "Action" strategy of player H) should be smaller than the other strategy. Otherwise, player H does not have to spend the cost of the governance protection. The number of nodes in the layer 0 for protecting $\eta$ blockchain networks in the layer 1 depends on the cost function and the optimal portion for the blockchain governance $\eta^1$ could be found as follows:

$$\eta^1 = inf\{\eta \geq 0 : \mathfrak{C}^1(\eta)\}, \quad (3.1)$$

where

$$\mathfrak{C}^1(\eta) = \sum_{l=0}^{\eta} \mathbb{E}\big[\mathfrak{C}_1^l(B_\eta)_{Total}\big]. \quad (3.2)$$

From (3.1), we have

$$\mathfrak{C}_{NoA}^l = V_l \cdot q_l^0, \tag{3.3}$$

$$\mathfrak{C}_{Act}^l(B) = c_1(B)\left(1 - q_l^1(B)\right) + (c_1(B) + V_l) \cdot q_l^1(B), \tag{3.4}$$

$$q_l^0 = \mathbb{E}\left[\mathbf{1}_{\left\{A_\nu^l \geq \frac{M_l}{2}\right\}}\right], \; \mathbb{E}[q_l^1(B)] = \mathbb{E}\left[\mathbb{E}\left[\mathbf{1}_{\left\{A_\nu^l \geq \frac{M_l}{2} + B\right\}}\Big|B\right]\right]. \tag{3.5}$$

We would like to design the enhanced blockchain network governance that can take the action at the decision making moment $\tau_{\nu-1}^l$. The governance in the blockchain is followed by the decision making parameter.

$$\begin{aligned}\mathfrak{C}_{\mathbf{1}}^l(B)_{Total} &= \mathfrak{C}_{Act}^l(B) \cdot \mathbb{E}\left[\mathbf{1}_{\left\{A_{\nu-1}^l < \frac{M_l}{2}\right\}}\right] + \mathfrak{C}_{NoA}^l \cdot \mathbb{E}\left[\mathbf{1}_{\left\{A_{\nu-1}^l \geq \frac{M_l}{2}\right\}}\right] \\ &= \{c^1(B)(1 - q_l^1(B)) + (c_1(B) + V_l) \cdot q_l^1(B)\}p_{A_{-1}} + V_l \cdot q^0(1 - p_{A_{-1}}) \\ &= \{c^1(B) + V_l \cdot q_l^1(B)\}p_{A_{-1}} + V_l \cdot q^0(1 - p_{A_{-1}}).\end{aligned} \tag{3.6}$$

where

$$p_{A_{-1}}^l = \mathbf{P}\left\{A_{\nu-1}^l < \frac{M_l}{2}\right\} = \sum_{k=0}^{\left\lfloor \frac{M_l}{2} \right\rfloor} \mathbf{P}\{A_{\nu-1}^l = k\}, \tag{3.7}$$

from (2.41),

$$P\{B_\eta = j\} = \binom{\eta}{j}(\rho^{\mathbf{1}})^j(1 - \rho^{\mathbf{1}})^{\eta-j}, \tag{3.8}$$

$$\rho^{\mathbf{1}} = \left(\frac{1}{\eta + 1}\right)\mathbb{E}\left[\sum_{k=0}^{\eta}\mathbf{1}_{\left\{H_\nu^k \geq \frac{M_k}{2}\right\}}\right], k = 0, \ldots, \eta. \tag{3.9}$$

From (3.2), we have

$$\mathbb{E}\left[\mathfrak{C}_{\mathbf{1}}^l(B_\eta)_{Total}\right] = \mathbb{E}\left[\mathfrak{C}_{Act}^l(B_\eta)\right] \cdot \mathbb{E}\left[\mathbf{1}_{\left\{A_{\nu-1}^l < \frac{M_l}{2}\right\}}\right] + \mathfrak{C}_{NoA}^l \cdot \mathbb{E}\left[\mathbf{1}_{\left\{A_{\nu-1}^l \geq \frac{M_l}{2}\right\}}\right].$$

For each blockchain network in the layer 1, we will not take any action for the $l$-th blockchain network until the time $\tau_{\nu^l-1}$ and it still have the chance that all nodes are governed by an attacker if the attacker catches more than the half of nodes at $\tau_{\nu-1}$ (i.e., $\{A_{\nu-1}^l \geq \frac{M_l}{2}\}$). If the attacker catches the less than half of all nodes at $\tau_{\nu-1}^l$ (i.e., $\{A_{\nu-1}^l < \frac{M_l}{2}\}$), then the defender could take the action to avoid the attack at $\tau_\nu^l$.

In the layer 0, the acceptance rate of a strategic alliance in a blockchain network $\alpha$ depends on the cost function and the optimal portion for the strategic alliance for blockchain governance $\alpha^{\mathbf{0}}$ (SABGG) could be found as follows:

$$\alpha^0 = \inf\{\alpha \geq 0 : \mathfrak{C}^0_{NoA}(r^0) \geq \mathfrak{C}^0_{Act}(\alpha)\}, \tag{3.10}$$

where (at the moment $t_{\nu-1}$),

$$\mathfrak{C}^0_{NoA}(r^0) = U_0 \cdot r^0, \tag{3.11}$$

$$\mathfrak{C}^0_{Act}(\alpha, \eta) = c^0_\eta(\alpha)(1 - r^1_\alpha) + (c^0_\eta(\alpha) + U_0)r^1_\alpha, \tag{3.12}$$

$$r^0 = \mathbb{E}\left[\mathbf{1}_{\{C_\nu \geq \frac{\eta}{2}\}}\right], \quad r^1_\alpha = \mathbb{E}\left[\mathbf{1}_{\{C_\nu \geq B_\eta\}}\right]. \tag{3.13}$$

From (3.9) and (3.10), the optimal acceptance rate for the layer 0 is determined as follows:

$$\alpha^* = \min\{\rho^1, \alpha^0\} \tag{3.14}$$

and the defender will not take any action until the time $t_{\nu-1}$ and there is the chance that all nodes are governed by the attacker if the attacker catches more than the half of nodes at $t_{\nu-1}$ (i.e., $C_{\nu-1} \geq \lceil \frac{N}{2} \rceil$). If the attacker catches the less than half of all nodes at $t_{\nu-1}$ (i.e., $C_{\nu-1} < \lceil \frac{N}{2} \rceil$), then the defender could take the action to avoid the attack at $t_\nu$. The total cost for developing the enhanced blockchain network is as follows:

$$\mathfrak{C}^0(\alpha, \eta)_{Total} = \mathfrak{C}^0_{Act}(B) \cdot \mathbb{E}\left[\mathbf{1}_{\{C_{\nu-1} < \frac{\eta}{2}\}}\right] + \mathfrak{C}^0_{NoA} \cdot \mathbb{E}\left[\mathbf{1}_{\{C_{\nu-1} \geq B\}}\right] \tag{3.15}$$
$$= \left\{c^0_{\{\alpha,\eta\}}(1 - r^1_\alpha) + \left(c^0_{\{\alpha,\eta\}} + U_0\right)r^1_\alpha\right\}p_{c_{-1}} + U_0 \cdot r^0(1 - p_{c_{-1}}),$$

where

$$p_{c_{-1}} = \mathbf{P}\left\{C_{\nu-1} < \left\lceil \frac{\eta}{2} \right\rceil\right\} = \sum_{k=0}^{\lfloor \frac{\eta}{2} \rfloor} \mathbf{P}\{C_{\nu-1} = k\}. \tag{3.16}$$

The optimal portion for the blockchain governance $\eta^1$ could be found as follows:

$$\eta^0 = \inf\{\eta \geq 0 : \mathfrak{C}^0(\eta)\}, \tag{3.17}$$

Because $\Theta_{\lceil \frac{\eta}{2} \rceil}(1, y_0, 1, 1, 1, 1)$ from (2.21) is the probability generating function of $C_{\nu-1}$, the probability mass could be found as follows:

$$\mathbf{P}\{C_{\nu-1} = k\} = \lim_{y_0 \to 0} \frac{1}{k!} \frac{\partial^k}{\partial y_0^k} \Theta_{\lceil \frac{N}{2} \rceil}(1, y_0, 1, 1, 1, 1), \quad k = 0, \ldots, \left\lfloor \frac{\eta}{2} \right\rfloor. \tag{3.18}$$

From (3.1) and (3.17), the optimal number of the reserved nodes for the layer 0 in the layer 1 could be found as follows:

$$\eta^* = \min\{\eta^1, \eta^0\}. \tag{3.19}$$

## 4. SIMULATION RESULTS

The safety mode is considered for protecting a network. Theoretically, the BGG based network takes a preliminary action to avoid a 51 percent attack by an attacker. The simulations in this section are targeted to find optimal values of the hyper parameters including an optimal number of reserved nodes in the layer 1 and the acceptance rate for the strategic alliance for the BGG based network in the layer 0.

### 4.1. Preliminaries

The safety mode is considered for protecting a multi-layered network. Theoretically, a BGG based network takes a preliminary action to avoid a 51 percent attack by an attacker. The action may actually happen before governing more than half of nodes by an attacker or after. Two points of the Proof-Of-Work (POW) are considered as action points: one is the moment that passes more than a half of nodes in the networks which are more than $\frac{M_l}{2}, l = 0, 1, \ldots \eta$ for the layer 1 and more than $\frac{\eta}{2}$ for the layer 0. The other is one step prior to pass more than half nodes in both layers. The best situation shall be that the safety mode is executed when an attacker takes more than a half of nodes but the network is protected by releasing additional backup nodes. But it is noted that attempting to govern more than a half of nodes may be happened even after exciting the safety mode.

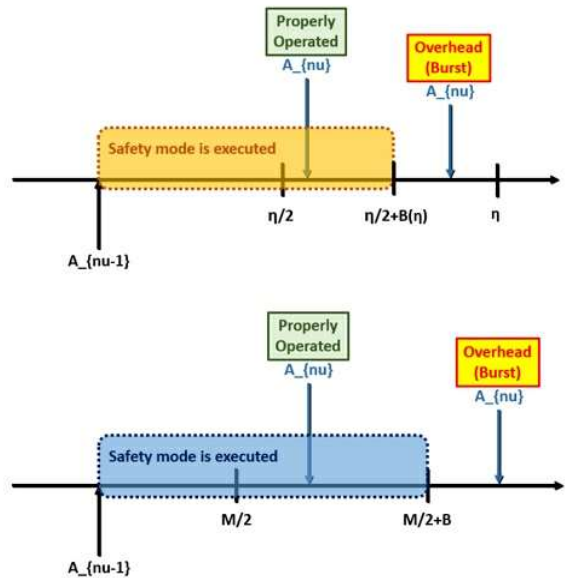

**Figure 4.** Operation of the safety mode of the layer 0 (above) and the layer 1 (bottom)

The simulation is processed by following strategies:
>  1) Two cases (one is with the BGG, the other is without the BGG) per each layer are simulated.
>  2) Simulating the 51 percent attack to evaluate whether the nodes in the network are protected by the BGG or not.
>  3) If the number of nodes governed by attack is more than a half at $\tau_\nu^l$ (i.e., $A_\nu^l \geq \frac{M_l}{2}$, $l = 0, 1, \ldots, \eta$), the networks in the layer 1 are burst.
>  4) If the number of nodes governed by attack is more than a half at $t_\nu$ (i.e., $C_\nu \geq \frac{\eta}{2}$), the network of the layer 0 is burst.
>  5) The safety modes for each layer are randomly executed based on the Binomial random variables.
>  6) The observation (i.e., the duration of the proof-of-work) are same within the same layer.

### 4.2. Optimizing backup nodes for the layer 1

This simulation considers that 41 (i.e., $\eta + 1$) IoT networks (as the layer 1) are hooked up as a single network (as the layer 0) and each IoT network has up to 40 backup nodes for security modes. The simulation goes 1000 trials and find the optimal number of backup nodes based on the cost efficiency. It has been executed 4 times with 1000 trials and the optimal values vary (see Figure 5).

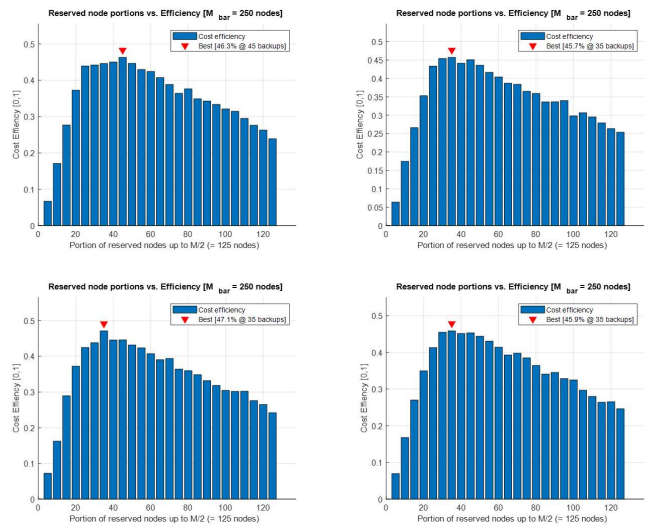

**Figure 5.** Simulation results to find the optimal value of backup nodes

It shows that the optimal number of backup nodes are ranged between 35 and 45 nodes. The cost efficiency is around 46 percents which indicates 250 BGG based IoT networks is 46 percents cheaper than atypical IoT networks in terms of overall operating costs.

### 4.3. Acceptance rate of strategic alliance in the layer 0
In the layer 0, the acceptance rate is a vital matter because the layer 0 network adapts the Strategic Alliance for Blockchain Governance Game (SABGG) [9]. This simulation robustly finds the optimal acceptance rate with correspond to number of nodes in the layer 0.

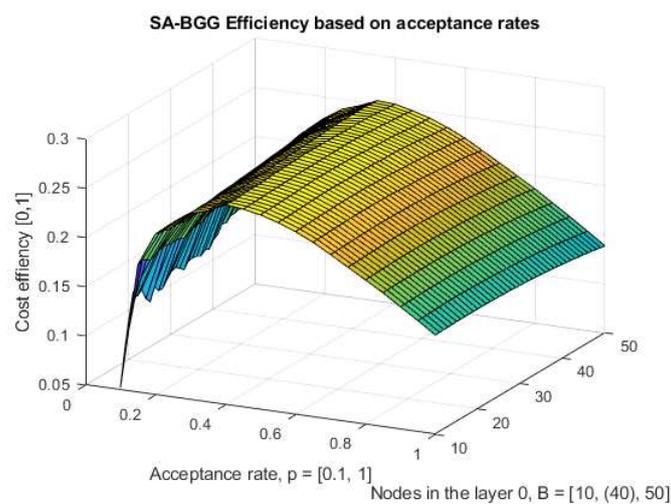

**Figure 6.** Acceptance rate optimization in the layer 0

The simulation shows that around 46 percents acceptance rate will give the best effort regardless of number of nodes in the layer 0 (see Figure 6).